\documentclass[letter,twocolumn]{jpsj2}
\title{Field Reentrance of the Hidden Order State of URu$_2$Si$_2$ under Pressure}

\author{%
Dai~\textsc{Aoki}\thanks{E-mail address: aokidai@gmail.com}, %
Fr\'ed\'eric~\textsc{Bourdarot},
Elena~\textsc{Hassinger},
Georg~\textsc{Knebel},
Atsushi~\textsc{Miyake},
St\'ephane~\textsc{Raymond},
Valentin~\textsc{Taufour}, and
Jacques~\textsc{Flouquet}
}

\inst{%
INAC/SPSMS, CEA-Grenoble, 17 rue des Martyrs, 38054 Grenoble, France\\
}
\recdate{\today}

\abst{%
Combination of neutron scattering and thermal expansion measurements under pressure
shows that the so-called hidden order phase of URu$_2$Si$_2$ 
reenters in magnetic field when antiferromagnetism (AF) collapses at $H_{\rm AF}(T)$.  Macroscopic pressure studies of the HO--AF boundaries were realized at different pressures 
via thermal expansion measurements under magnetic field using a strain gauge.
Microscopic proof at a given pressure is the reappearance of the resonance at $\mbox{\boldmath $Q_0$}=(1,0,0)$ under field 
which is correlated with the collapse of the AF Bragg reflections at $\mbox{\boldmath $Q_0$}$. 
}

\kword{URu$_2$Si$_2$, heavy fermion, hidden order, phase transition, magnetism, electronic structure}

\begin{document}
\maketitle

Pressure $(P)$ and magnetic field $(H)$ are nice tools to tune correlated systems through a critical point of different ground states.~\cite{Flo06_review}
Here, we use a combination of these two variables in order to clarify the so-called hidden order (HO) of the heavy fermion compound URu$_2$Si$_2$. 
Pressure studies have shown that the system jumps from the HO phase to an antiferromagnetic (AF) phase 
at a pressure $P_{\rm x}\simeq 0.5\,{\rm GPa}$ at $0\,{\rm K}$. 
In zero magnetic field the HO, AF and paramagnetic (PM) phases form three distinct phases and the transition lines between these phases seems to meet in a triple point with $P_{\rm c}\sim 1.2\,{\rm GPa}$ and $T_0 = T_{\rm N} \sim 18.5\,{\rm K}$.~\cite{Ami07,Cha02,Has08}
$T_0$, $T_{\rm N}$ and $T_{\rm x}$ are the critical temperatures for the PM--HO, PM--AF and HO--AF phase transitions, respectively.
Selecting a pressure $P$ in the pressure window between $P_{\rm x}$ and $P_{\rm c}$, it was observed
in a recent neutron scattering experiment~\cite{Vil08}  that the HO phase between $T_0$ and $T_{\rm x}$
is characterized by a resonance at an energy of $E_0 \sim 2\,{\rm meV}$
for the wave vector $\mbox{\boldmath $Q_0$}=(1,0,0)$ as already reported at ambient pressure.~\cite{Bro87,Bou03,Wie07}
This resonance corresponds to  large longitudinal magnetic fluctuations at the vector $\mbox{\boldmath $Q_{\rm AF}$} = (0,0,1)$. Remarkably, the resonance at $\mbox{\boldmath $Q_0$}$ 
collapses below $T_{\rm x}$, while a large elastic magnetic Bragg peak appears at the same wave vector $\mbox{\boldmath $Q_0$}$ which corresponds to the antiferromagnetic phase with the propagation vector $\mbox{\boldmath $Q_{\rm AF}$}=(0,0,1)$.~\cite{Ami07}

The onset of AF at $T_{\rm x}$ in the pressure range between $P_{\rm x}$ and $P_{\rm c}$, and at $T_{\rm N}$ above $P_{\rm c}$ 
is associated with a change of symmetry as two different U magnetic sites appear now in the primitive body centered tetragonal (bct) unit cell. Thus the lattice is no more bct but tetragonal.
There are evidences that such a change in symmetry may occur already in the HO phase: 
i) at $T_0$ and $T_{\rm N}$ the transition from PM to the ordered phase
seems to be associated with a similar Fermi surface reconstruction, i.~e.~
a large drop of the electronic carrier density due to a partial gap opening $\Delta_{\rm B}$
in the band structure in consequence of the change from bct to tetragonal crystal structure,~\cite{Map86,Sch87,Beh05} 
and 
ii) at very low temperatures no change of the Fermi surface can be detected through $P_{\rm x}$, 
at least in de Haas-van Alphen (dHvA) branch $\alpha$.~\cite{Nak03}
The $P$ invariance of these two observations suggests that the transition from HO to AF is isostructural for the lattice.
Due to the Ising character of the U magnetism in URu$_2$Si$_2$, 
the onset of AF can preserve the same tetragonal lattice as in the HO phase.

Band structure calculations with either bct or tetragonal crystal structure (assuming the 5\textit{f} electrons of U atoms to be itinerant)
show that the system is a compensated metal for both assumptions,~\cite{Yam00_URu2Si2,Har09,Elg08} in agreement with experimental observations.~\cite{Ohk99}
Furthermore the transition from bct to tetragonal symmetry in the crystal structure 
leads to a decrease of carrier number by a factor between 3 and 5.
Recently, the partial band gap $\Delta_{\rm B}$ in the AF phase was calculated as a function 
of the strength of the sublattice magnetization $M_0$.~\cite{Elg08}
On the basis of experimental findings,
the authors proposed that a finite band gap $\Delta_{\rm B}$ survives in the HO state
due to the strong longitudinal magnetic fluctuations at {\boldmath $Q_0$},
which are the origin of the observed resonance.

Here we present a new route to study the interplay of HO and AF phases by tuning the magnetic field under high pressure.
The main intention is to show the interplay between the field $H_{\rm M}(T,P)$, 
characteristic of a complete collapse of the band gap, 
and the magnetic field $H_{\rm AF}(T,P)$ which defines the AF boundary with other phases (HO or PM).
At low pressure the HO state is a robust ground state at least up to $H\sim 34\,{\rm T}$ for the field along $c$-axis. 
Only above $H_{\rm M}\simeq 40\,{\rm T}$,
a polarized paramagnetic state is achieved through a metamagnetic transition and
a high number of electronic carrier is restored.~\cite{Oh07,Lev09}
Just below $H_{\rm M}$, in the field window $34$--$40\,{\rm T}$, a cascade of different phases has been observed.~\cite{Jo07}
Their identification  is not yet completed. 
Furthermore we investigate if the pressure induced AF phase disappears at a field $H_{\rm AF} (T=0)$ lower than $H_{\rm M} (T=0)$,
furthermore if the HO state reappears when AF is suppressed.
To study the respective ($H,T$) boundary between the HO and AF phases under pressure,
two different approaches, namely neutron scattering and thermal expansion measurements were realized.

Several high-quality single crystals of URu$_2$Si$_2$ were grown by the Czochralski method in a tetra-arc furnace.
The samples were cut by a spark cutter 
and successively annealed at $1075\,^\circ{\rm C}$ for 5 days under ultra high vacuum of $10^{-10}\,{\rm torr}$.
Thereafter the samples were checked by X-ray Laue photograph and resistivity measurements. 
The residual resistivity ratio (RRR) is higher than $100$, 
indicating the high quality of the samples. 

In the continuation of the previous neutron scattering study,~\cite{Vil08} 
similar experiments were performed in magnetic field up to $H=15\,{\rm T}$ 
at a constant pressure in the window $P_{\rm x} < P < P_{\rm c}$,
This microscopic probe will show that at $P=0.72$~GPa above $H_{\rm AF}(0)\sim 12\,{\rm T}$,
the HO state is restored with its characteristic $\mbox{\boldmath $Q_0$}=(1,0,0)$ resonance.
The neutron experiment under pressure were carried out 
by using a CuBe pressure cell on the cold-triple axis spectrometer IN14 at ILL. 
On a large single crystal ($6\,{\rm mm}$ in diameter and $11\,{\rm mm}$ in length along $c$-axis) a strain gauge was glued on top of the sample and then put into the cell.  As pressure transmitting medium deserved a mixture of Fluorinert FC84/FC87 (1:1). 
The measurements were performed in a $^4$He cryostat and a magnetic field up to $15\,{\rm T}$ has been applied along the $c$-axis. The strain gauge acts as indicator for the HO--AF phase transition at fixed pressure. 
We determined $T_{\rm x} = 10\,{\rm K}$ and $T_{\rm 0} = 18.5\,{\rm K}$ at zero field at $0.72\,{\rm GPa}$.

The necessity to use a large crystal in the inelastic neutron scattering experiment is at expense of performing the measurement under ideal homogeneous pressure conditions. However, this difficulty is bypassed by the performance of the thermal expansion measurements on a small crystal by a strain gauge method.~\cite{Mot03}
The $(H,T)$ phase diagrams at different fixed pressures have been determined by thermal expansion measurements using a strain gauge in a CuBe-NiCrAl hybrid pressure cell with Daphne oil 7373  as a pressure medium. 
The strain gauge was glued on the $c$-plane of the sample with dimensions of $2\times 2\times 0.5\,{\rm mm}^3$.
The magnetic field was applied along $c$-axis.
Measurements up to $9\,{\rm T}$ have been performed in a commercial Quantum Design PPMS, for the high field measurements up to $16\,{\rm T}$ a home-made cryostat has been used.
This technique allows us to present a direct determination of the $H_{\rm AF}(T)$, $H_{\rm M}(T)$ boundaries at a fixed pressure.
These measurements confirm the reentrance of the HO phase under magnetic field higher than $H_{\rm AF}$ for pressures above $P_{\rm x}$ and furthermore indicate that $H_{\rm AF}(T=0)$ increases strongly with increasing pressure.
Above $P_{\rm c}$ at $P \sim 2\,{\rm GPa}$, the reentrance of the HO phase cannot be detected for $H$ up to $16\,{\rm T}$.  
In both experiments the pressure was determined by measuring the superconducting transition of Pb by ac susceptibility.

\begin{figure}[htb]
\begin{center}
\includegraphics[width=0.6 \hsize,clip]{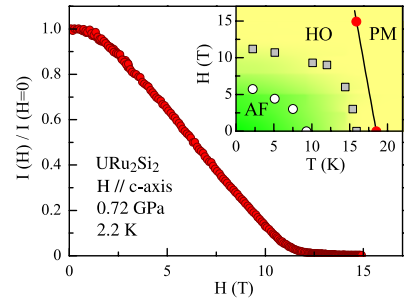}
\end{center}
\caption{(Color online) At $P=0.72\,{\rm GPa}$, field dependence of the intensity $I_{\rm M}(H)$ of the $\mbox{\boldmath $Q_0$}=(1,0,0)$
magnetic Bragg reflection measured at low temperature ($T=2.2\,{\rm K}$). 
The inset shows $(H,T)$ phase diagram.
The AF--HO boundaries as defined from the onset of the intensity (squares) or at $50\,{\%}$ of $I_{\rm M}(0)$ (open circles). 
Solid circles indicate $T_0$.}
\label{fig:neutron_elastic}
\end{figure}
\begin{figure}[htb]
\begin{center}
\includegraphics[width=0.6 \hsize,clip]{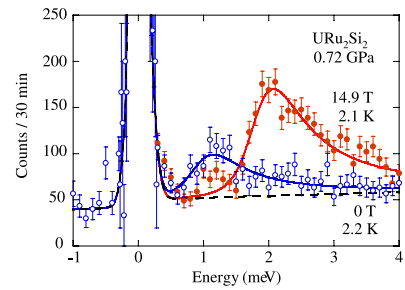}
\end{center}
\caption{(Color online) At $P=0.72\,{\rm GPa}$, inelastic intensity at $\mbox{\boldmath $Q_0$}=(1,0,0)$
measured at zero field with $90\,{\%}$ of AF phase (open circles)
and at $14.9\,{\rm T}$ far above $H_{\rm AF}$ (solid circles). A dashed line is the background.}
\label{fig:neutron_inelastic}
\end{figure}
Figure~\ref{fig:neutron_elastic} represents the field variation of the elastic magnetic neutron intensity $I_{\rm M}(H)$ at $T=2.2$~K and $P = 0.72$~GPa
measured for the wave vector $\mbox{\boldmath $Q_0$}=(1,0,0)$.
Due to (i) the proximity of $P_{\rm x}$, (ii) the inhomogeneity produced by the Fluorinert pressure transmitting medium
(presumably during its solidification on too fast cooling, it is known that a large pressure gradient parallel
to the load can appear on solidifying Fluorinert)~\cite{Koy07}
and (iii) the necessity to use a large single crystal which covers $70\,{\%}$ of the pressure chamber,
the elastic signal $I_{\rm M} (T,H)$ measured at different temperatures and magnetic fields
may not correspond to the one observed in perfect hydrostatic conditions. 
Thus depending on the criterion chosen for the determination of the phase transition
from AF to HO (either the extrapolated full loss of the elastic signal or $50\,{\%}$ of the full signal,
the $H_{\rm AF}(T)$ borderline at $P= 0.72$~GPa can be drawn as shown in the inset of Fig.~\ref{fig:neutron_elastic}.
Under these (not ideal) experimental pressure conditions we can estimate that
$10\,{\%}$ of HO phase persists in the dominant $90\,{\%}$ AF ground state at this pressure.
However, the spectacular new result is that far above $H_{\rm AF}$ at $14.9\,{\rm T}$,
the inelastic response at $\mbox{\boldmath $Q_0$}=(1,0,0)$ characteristic of the HO phase 
reappears as shown in Fig.~\ref{fig:neutron_inelastic}.
The remaining response signal at zero field is only the consequence of the $10\,{\%}$ residual HO fraction.
For the AF fraction, it has been confirmed that the resonance disappears at zero field.~\cite{Vil08}
A more detailed discussion of the neutron scattering spectra obtained at various temperatures and different magnetic fields including scans in PM, HO and AF phases will be published in an subsequent extended paper~\cite{Bou09} 
since here we focus on the pressure dependence of the $H_{\rm AF}$ boundary.

\begin{figure}[htb]
\begin{center}
\includegraphics[width=0.8 \hsize,clip]{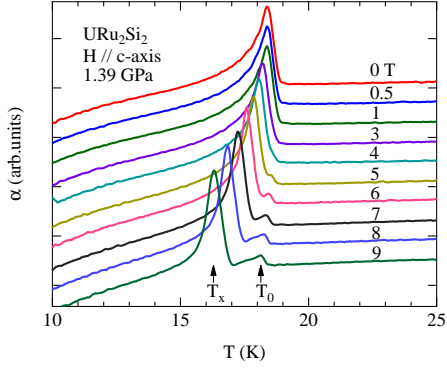}
\end{center}
\caption{(Color online) Thermal expansion measurements as a function of $T$ for different magnetic fields from $0$ to $9\,{\rm T}$ at $P=1.39\,{\rm GPa}$ (data are shifted for clarity). 
The arrows indicate $T_{\rm x}$ and $T_{\rm 0}$ at $9\,{\rm T}$. }
\label{fig:Texp}
\end{figure}

\begin{figure}[htb]
\begin{center}
\includegraphics[width=0.8 \hsize,clip]{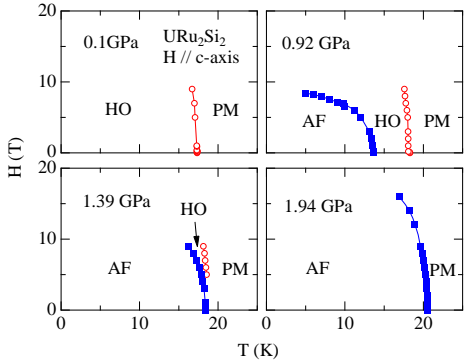}
\end{center}
\caption{(Color online) $(H,T)$ phase diagram at $P=0.1$, $0.92$, $1.39$ and $1.94\,{\rm GPa}$. Open circles and solid squares denote $H_{\rm M}(T)$ and $H_{\rm AF}(T)$, respectively.}
\label{fig:HT_phase}
\end{figure}

Figure~\ref{fig:Texp} represents the temperature dependence of the thermal expansion coefficient $\alpha$ for different fields at $P=1.39\,{\rm GPa}$
measured by using the strain gauge.
The location of $T_{\rm x} (H)$ and $T_{\rm 0}(H)$ is defined by the maxima of $\alpha$.
In Fig.~\ref{fig:HT_phase} we have plotted the $H_{\rm AF}(T)$, $H_{\rm M}(T)$ boundaries for different pressures,  
$P=0.1$, $0.92$, $1.39$, and $1.94$~GPa .
At $P=0.1\,{\rm GPa} < P_{\rm x}$, only the HO boundary is observed. 
For $P_{\rm x} < P=0.92\,{\rm GPa} < P_{\rm c}$, 
a distinct separation exists  between the AF and HO boundaries.
Just near $P_{\rm c}$ at $1.39\,{\rm GPa}$, the magnetic field seems to lift the degeneracy between HO and AF phase boundary,
while above $P_{\rm c}$ at $P=1.94\,{\rm GPa}$ up to the highest applied field of $16\,{\rm T}$,
only the AF phase can be observed. However, further experiments with the determination of the volume (here we measured only one direction) have to be performed to rule out the possibility of a triple point in the $(H, T)$ plane at fixed pressure above $P_{\rm c}$ under magnetic fields.

\begin{figure}[htb]
\begin{center}
\includegraphics[width=0.7 \hsize,clip]{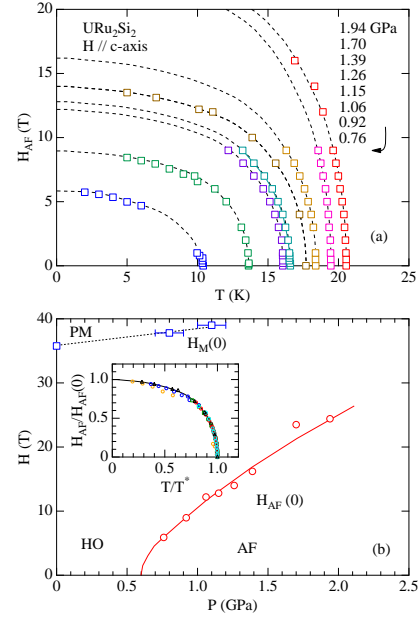}
\end{center}
\caption{(Color online) (a) Temperature dependence of $H_{\rm AF}$ at different pressure. 
(b) Pressure dependence of $H_{\rm AF}(0)$ extrapolated to $0\,{\rm K}$ (open circles),
which is obtained from the scaling reduction (inset), 
where $T^\ast$ is $T_{\rm x}$ for $P < P_{\rm c}$ or 
                  $T_{\rm N}$ for $P > P_{\rm c}$.
Open squares in panel (b) give $H_{\rm M}(0)$ as function of pressure from ref.~\citen{Jo07}.
Dashed lines in panel (a) are the results of scaling.}
\label{fig:H_AF}
\end{figure}

Figure~\ref{fig:H_AF}(a) shows $H_{\rm AF} (T)$ at various pressures.  
The extrapolation of $H_{\rm AF} (T)$ to $0\,{\rm K}$ is obtained by scaling
$H_{\rm AF}(T)/H_{\rm AF}(0)$ as a function of $T/T^\ast$ for different pressures, where $T^\ast$ is $T_{\rm x}$ (for $P< P_{\rm c}$) or $T_{\rm N}$ (for $P > P_{\rm c}$).   
$H_{\rm AF}(0)$ increases strongly with pressure as shown in Fig.~\ref{fig:H_AF}(b).
The concomitant strong pressure dependence of $T_{\rm x}$ and $H_{\rm AF}(0)$ explains 
the broadening of the AF boundary in the inhomogeneous $P$ neutron scattering experiment (see above).
By contrast as indicated in Fig.~\ref{fig:H_AF}(b), $H_{\rm M}(0)$ is only weakly $P$ dependent.~\cite{Jo07}
The splitting between $H_{\rm AF}(T, P)$ and $H_{\rm M}(T,P)$ for $P_{\rm x} < P < P_{\rm c}$ observed here 
is similar to the effects reported in the Rh doped compound
U(Ru$_{0.98}$Rh$_{0.02}$)$_2$Si$_2$ by specific heat, magnetization and neutron scattering.~\cite{Bou05}
However, an additional difficulty in this case is that the doping changes significantly the carrier concentration
and thus may spoil the bare phenomena. 
\begin{figure}[htb]
\begin{center}
\includegraphics[width=0.7 \hsize,clip]{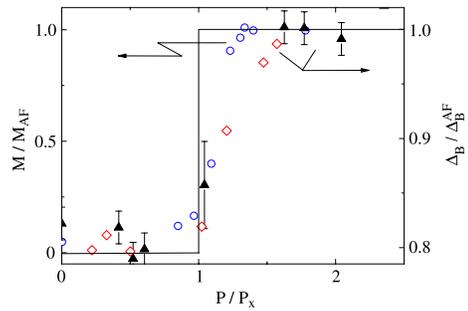}
\end{center}
\caption{(Color online) Band gap $\Delta_{\rm B}$ normalized to the size of the gap in the antiferromagnetic state $\Delta_{\rm B}^{\rm AF}$ above $P_{\rm c}$ derived from the resistivity as a function of $P/P_{\rm x}$. Solid triangles are from ref.~\citen{Has08}, open diamonds from ref.~\citen{Jef08}. 
By comparison, predicted variation of the sublattice magnetization in ideal hydrostatic condition (solid line)
and the normalized sublattice magnetization (open circles) from ref.~\protect\citen{Ami07}.}
\label{fig:6}
\end{figure}

Having these new results in mind, it is interesting to reanalyze previous results at $P=0$
on the field dependence of the intensity ($I_{\rm M} \sim M^2$) of the tiny sublattice magnetization $(M)$ ($M_0\sim 0.03\,\mu_{\rm B}$ at $T=0$~K)
detected in the HO phase at ambient pressure.~\cite{Bou03}
As observed here in the pressure window $P_{\rm x} < P < P_{\rm c}$,
the magnetic field response of $I_{\rm M}(H)$ for a fixed pressure and at low temperature through the $H_{\rm AF}(T)$ boundary is smeared out and
quite different from the $H_{\rm M}(T)$ boundary defined by the thermal expansion measurements.~\cite{}
Assuming that the tiny sublattice magnetization is a mark of residual components of the AF phase,
the long tail observed in $I_{\rm M} (H)$ in the neutron scattering experiment points out 
a large distribution in the AF parameters ($T_{\rm x}$, $T_{\rm N}$, $H_{\rm AF}$)
due to local pressure inhomogeneities which may result in residual AF fractions.
It is worthwhile to note that in the same way that the superconducting transition
observed by resistivity far above $P_{\rm x}$~\cite{Has08,Bou05,Jef08}
(whereas homogeneous superconductivity has been observed to collapse at $P_{\rm x}$)~\cite{Ami07,Has08} appears as a
consequence of a surviving fraction of the HO phases inside the AF phase.
The sensitivity to ideal $P$ conditions can be explained qualitatively by 
(i) the small value of $P_{\rm x}$,
(ii) the strong anisotropy and  
(iii) the possibility to get locally pressure inhomogeneities near imperfection comparable to $P_{\rm x}$ or even $P_{\rm c}$.
However, these inhomogeneous tails are not easy to explain quantitatively.
An open delicate question is whether even for a perfect crystal 
still finite specific correlated nano-structures will not exist.

Recently the link between the band gap  $\Delta_{\rm B}$ and of the sublattice magnetization $M_0$ has been discussed.~\cite{Elg08}
In agreement with previous works,~\cite{Sch93,Jef08} the pressure dependence of $\Delta_{\rm B}$ 
is evaluated from the drop of the resistivity at low temperatures 
by fitting $\rho (T) = \rho_0 + A T^x + B T/\Delta_{\rm B} ( 1 + 2T/\Delta_{\rm B} ) \exp (-\Delta_{\rm B}/T)$. 
The free exponent $x$ in the $AT^x$ term takes into account the experimental fact that we do not find a Fermi liquid $T^2$ temperature dependence at low temperature above the superconducting transition. 
Rigorously, the last term describes  the scattering of conduction electrons with magnons with a gap $\Delta_{\rm B}$ in the excitation spectra.  Therefore the estimation of $\Delta_{\rm B}$ by this formula can only be qualitative for URu$_2$Si$_2$ since the scattering at low temperature below $T_0$ is dominated by the longitudinal fluctuations as observed in inelastic neutron scattering experiments. 
Thus the link between $\Delta_{\rm B}$ derived by transport and the inelastic neutron spectrum is not obvious. 
As indicated in Fig.~\ref{fig:6} in agreement with previous studies~\cite{Sch93,Jef08}
$\Delta_{\rm B}$ increases through $P_{\rm x}$, 
the pressure where the jump of sublattice magnetization occurs from $M_0 \sim 0$ to $0.4\,\mu_{\rm B}$.
As a magnetic field weakens $\Delta_{\rm B}$,~\cite{Men96,van97} 
the simple idea is that in the AF phase under magnetic field the gap $\Delta_{\rm B}$ may decrease 
below a critical value $\Delta_{\rm B}^0$ where the system will jump from AF to HO phase.

Up to now, there is no direct evidence of the bct to tetragonal transition at $T_0$ in the HO phase.
At first glance considering the magnetism, the HO phase appears as PM, however, an orbital ordering may be hidden.~\cite{Cha02}
A favorable factor therefore may be the initial mixed valence character in its PM state.~\cite{Has04}
The possibility of fancy orbital ordering may come from the quasi-degeneracy between the trivalent and tetravalent uranium configuration;
thus a tiny change of volume will favor the electronic system to be normalized to one or another configuration (see for example, TmSe in ref.~\citen{Der06}).

These new experiments reveal the reentrance of the HO phase under magnetic field for pressures $P > P_{\rm x}$  when the AF order is suppressed.
A microscopic signature is the reemergence of the resonance at $\mbox{\boldmath $Q_0$}$ which is
associated to the collapse of the sublattice magnetization. 
This effect is strongly related to the Ising type case 
of URu$_2$Si$_2$ with strong valence fluctuations in the PM state.
It is proposed that the order parameter of the HO state is strongly connected to 
the amplitude of the band gap $\Delta_{\rm B}$. 
The band gap might appear through symmetry change of the lattice from bct to tetragonal. 
This symmetry change is proved to occur in the AF phase 
and suggested to persist also at the HO phase transition 

We thank T. Komatsubara and M. Suzuki for their technical support of crystal growth.
This work was financially supported by French ANR projects ECCE and CORMAT.



\begin{thebibliography}{10}
\expandafter\ifx\csname url\endcsname\relax
  \def\url#1{\texttt{#1}}\fi
\expandafter\ifx\csname urlprefix\endcsname\relax\def\urlprefix{URL }\fi

\bibitem{Flo06_review}
see e.g. J.~Flouquet: {\em Progress in Low Temperature Physics}, ed. W.~P. Halperin
  (Elsevier, Amsterdam, 2006) p.~139.

\bibitem{Ami07}
H.~Amitsuka, K.~Matsuda, I.~Kawasaki, K.~Tenya, M.~Yokoyama, C.~Sekine,
  N.~Tateiwa, T.~C. Kobayashi, S.~Kawarazaki, and H.~Yoshizawa: J. Magn. Magn.
  Mater. {\bf 310}  (2007) 214.

\bibitem{Cha02}
P.~Chandra, P.~Coleman, J.~A. Mydosh, and V.~T. V.: Nature {\bf 417} (2002) 831.

\bibitem{Has08}
E.~Hassinger, G.~Knebel, K.~Izawa, P.~Lejay, B.~Salce, and J.~Flouquet: Phys.
  Rev. B {\bf 77} (2008) 115117.

\bibitem{Vil08}
A.~Villaume, F.~Bourdarot, E.~Hassinger, S.~Raymond, V.~Taufour, D.~Aoki, and
  J.~Flouquet: Phys. Rev. B {\bf 78} (2008) 012504.

\bibitem{Bro87}
C.~Broholm, J.~K. Kjems, W.~J.~L. Buyers, P.~Matthews, T.~T.~M. Palstra, A.~A.
  Menovsky, and J.~A. Mydosh: Phys. Rev. Lett. {\bf 58} (1987) 1467.

\bibitem{Bou03}
F.~Bourdarot, B.~F\aa{}k, K.~Habicht, and K.~Proke{\v{s}}: Phys. Rev. Lett.
  {\bf 90} (2003) 067203.

\bibitem{Wie07}
C.~R. Wiebe, J.~A. Janik, G.~J. Macdougall, G.~M. Luke, J.~D. Garrett, H.~D.
  Zhou, Y.~J. Jo, L.~Balicas, Y.~Qiu, J.~R.~D. Copley, Z.~Yamani, and W.~J.~L.
  Buyers: Nature Physics {\bf 3} (2007) 96.

\bibitem{Map86}
M.~B. Maple, J.~W. Chen, Y.~Dalichaouch, T.~Kohara, C.~Rossel, and M.~S.
  Torikachvili: Phys. Rev. Lett. {\bf 56} (1986) 185.

\bibitem{Sch87}
J.~Schoenes, C.~Sch{\"o}nenberger, J.~J.~M. Franse, and A.~A. Menovsky: Phys.
  Rev. B {\bf 35} (1987) 5375.

\bibitem{Beh05}
K.~Behnia, R.~Bel, Y.~Kasahara, Y.~Nakajima, H.~Jin, H.~Aubin, K.~Izawa,
  Y.~Matsuda, J.~Flouquet, Y.~Haga, Y.~{\=O}nuki, and P.~Lejay: Phys. Rev. Lett.
  {\bf 94} (2005) 156405.

\bibitem{Nak03}
M.~Nakashima, H.~Ohkuni, Y.~Inada, R.~Settai, Y.~Haga, E.~Yamamoto, and
  Y.~\={O}nuki: J. Phys.: Condens. Matter {\bf 15} (2003) S2011.

\bibitem{Yam00_URu2Si2}
H.~Yamagami and N.~Hamada: Physica B {\bf 284-288} (2000) 1295.

\bibitem{Har09}
{H. Harima: private communication}.

\bibitem{Elg08}
S.~Elgazzar, J.~Rusz, M.~Amft, P.~M. Oppeneer, and J.~A. Mydosh:
arXiv:0809.2887.
  
\bibitem{Ohk99} H. Ohkuni, Y. Inada, Y. Tokiwa, K. Sakurai, R. Settai, T. Honma, Y. Haga, E. Yamamoto, Y. \={O}nuki, H. Yamagami, S. Takahashi, and T. Yanagisawa: Philos. Mag. B {\bf 79} (1999) 1045.

\bibitem{Oh07}
Y.~S. Oh, K.~H. Kim, P.~A. Sharma, N.~Harrison, H.~Amitsuka, and J.~A. Mydosh:
  Phys. Rev. Lett. {\bf 98} (2007) 016401.

\bibitem{Lev09}
J.~Levallois, K.~Behnia, J.~Flouquet, P.~Lejay, and C.~Proust: EPL {\bf 85}
  (2009) 27003.

\bibitem{Jo07}
Y.~J. Jo, L.~Balicas, C.~Capan, K.~Behnia, P.~Lejay, J.~Flouquet, J.~A. Mydosh,
  and P.~Schlottmann: Phys. Rev. Lett. {\bf 98} (2007) 166404.

\bibitem{Mot03}
G.~Motoyama, T.~Nishioka, and N.~K. Sato: Phys. Rev. Lett. {\bf 90} (2003)
  166402.
  
\bibitem{Koy07}
K.~Koyama-Nakazawa, M.~Koeda, M.~Hedo, and Y.~Uwatoko: Rev. Sci. Instr. {\bf 78} (2007) 066109.
  
\bibitem{Bou09}
F.~Bourdarot, et al. to be published.
  
\bibitem{Bou05}
F.~Bourdarot, B.~F{\aa}k, F.~Lapierre, I.~Sheikin, and P.~Lejay: Physica B {\bf
  359-361} (2005) 1132.

\bibitem{Jef08}
J.~R. Jeffries, N.~P. Butch, B.~T. Yukich, and M.~B. Maple: J. Phys.: Condens.
  Matter {\bf 20} (2008) 095225.

\bibitem{Sch93}
L.~Schmitt: PhD Thesis in Grenoble.

\bibitem{Men96}
S.~A.~M. Mentink, T.~E. Mason, S.~S{\"u}llow, G.~J. Nieuwenhuys, A.~A.
  Menovsky, J.~A. Mydosh, and J.~A. A.~J. Perenboom: Phys. Rev. B {\bf 53}
  (1996) R6014.

\bibitem{van97}
N.~H. {van Dijk}, F.~Bourdarot, J.~C.~P. Klaasse, I.~H. Hagmusa, E.~{Br\"{u}ck},
  and A.~A. Menovsky: Phys. Rev. B {\bf 56} (1997) 14493.

\bibitem{Has04}
E.~Hassinger, J.~Derr, J.~Levallois, D.~Aoki, K.~Behnia, F.~Bourdarot,
  G.~Knebel, C.~Proust, and J.~Flouquet: J. Phys. Soc. Jpn. Suppl. A {\bf 77}
  (2008) 172.

\bibitem{Der06}
J.~Derr, G.~Knebel, G.~Lapertot, B.~Salce, M.~M\'{e}asson, and J.~Flouquet: J.
  Phys.: Condens. Matter {\bf 18} (2006) 2089.

\end{thebibliography}

\end{document}